# Non-IID Recommender Systems: A Review and Framework of Recommendation Paradigm Shifting

Longbing Cao

*Advanced Analytics Institute, University of Technology Sydney, Sydney, NSW 2007, Australia*

A B S T R A C T

While recommendation plays an increasingly critical role in our living, study, work, and entertainment, the recommendations we receive are often for irrelevant, duplicate, or uninteresting products and services. A critical reason for such bad recommendations lies in the intrinsic assumption that recommended users and items are independent and identically distributed (IID) in existing theories and systems. Another phenomenon is that, while tremendous efforts have been made to model specific aspects of users or items, the overall user and item characteristics and their non-IIDness have been overlooked. In this paper, the non-IID nature and characteristics of recommendation are discussed, followed by the non-IID theoretical framework in order to build a deep and comprehensive understanding of the intrinsic nature of recommendation problems, from the perspective of both couplings and heterogeneity. This non-IID recommendation research triggers the paradigm shift from IID to non-IID recommendation research and can hopefully deliver informed, relevant, personalized, and actionable recommendations. It creates exciting new directions and fundamental solutions to address various complexities including cold-start, sparse data-based, cross-domain, group-based, and shilling attack-related issues.

## 1. Introduction

Recommendation [1,2] is a major application of big data [3,4]. It plays an increasingly important role in both core business and new economy, particularly when it involves social media, mobile services, online business, and study and living. In recent years, recommendation research has attracted significant attention in many communities, including recommender systems, information retrieval, social media, social networks, machine learning, data mining, and data engineering.

A quality recommendation service should refer the most relevant products (services or other items) to the right people at the right time. Intensive efforts have been made, especially in the recommendation and information-retrieval communities, to improve recommendation quality, by considering specific factors such as social relationships, friendship, user comments on purchased products, grouping similar behaviors or categories of products, and recommending products from another domain.

In most cases, however, we have seen irrelevant or even brand-damaging recommendations of products or services to us through channels including news portals, online shopping websites, and mobile applications. For example, a famous search engine website placed an advertisement suggesting a visit to Greek beaches alongside a news article about civil protest taking place in Greece. Another website recommended different kinds of fruit to a user who showed interest in kiwis (a New Zealand bird), assuming that the user's interest was in kiwi fruit. Online bookselling websites often list books that either duplicate those we have already purchased or are totally irrelevant.

Some critical questions facing the recommendation commu-

*E-mail address:* longbing.cao@gmail.com

nity include: Why are we recommended irrelevant or duplicated products and services? and more critically, What makes next-generation recommendation? To answer these questions, even more fundamental problems need to be studied, including:

- What foundational aspects have been missing in existing recommendation theories and systems that result in poor recommendations?
- How can informed, relevant, personalized, and actionable recommendations be made?
- How can recommendation quality be improved so that only those products that are relevant to individual or group user interests, preferences, and circumstances are pushed forward?
- What are new recommendation methodologies are essential for a unified theoretical framework that can capture the intrinsic characteristics and complexities in recommendation?
- What is the paradigm shift of recommendation research that fundamentally enables the next-generation research?
- What forms the foundation of next-generation recommendation?
- What are the new directions for the next generation of recommendation theories and systems?
- What new recommendation theoretical foundation can address typical challenges including cold-start, sparsity, cross-domain, group recommendation, and shilling attack?

While there are many aspects to be explored in order to address the above foundational problems, one particular perspective that is of interest in this paper concerns the in-depth understanding of recommended users and items, and the tight connection between the ratings given by a user to an item and the characteristics of users and items. This involves an in-depth understanding of the intrinsic characteristics and complexities in recommendation, and the nature of recommendation; that is, the heterogeneity and couplings (namely non-IIDness [5,6]) of ratings, user properties, and item properties, and the heterogeneity and couplings among these three aspects.

In existing recommendation research, various efforts have been made on high-level aspects such as user ratings on items, user social relationships, and comments on items. Such efforts may be generally categorized into the following aspects: ① estimating future ratings based on existing ratings, ② incorporating user comments on items into modeling, ③ incorporating user friendliness into modeling, ④ modeling group preferences or behaviors, and ⑤ learning the user preference transfer across domains. User behaviors of viewing or commenting on items are also modeled in Refs. [7,8]. More recently, coupling relationships between items and user groups have been modeled for recommendation [9–11], bringing low-level driving forces into rating dynamics estimation.

However, state-of-the-art recommendation research [2] has been built on the assumption that users, products, and ratings are independent and identically distributed (IID), resulting in IID models and methods [5]. No work has considered very low-level non-IID information about specific users and items. In this way, the fundamental driving forces of ratings are simplified or overlooked, which this author believes is a critical reason for the poor performance of existing recommender systems and services. For example, matrix factorization (MF) is a generic mathematical tool widely used in recommendation modeling. However, it will generate similar outcomes for houses and cars if the low-level properties of houses and cars are not involved and if houses and cars are treated as IID. This creates a significant gap between general high-level models and the specific low-level information associated with recommended users and items.

This paper focuses on discussing the driving role of such information in capturing the nature of recommendation and improving recommendation quality. By extending the brief discussion in Ref. [12] about non-IID recommendation theories and systems, a systematic framework and an in-depth understanding of recommendation nature are provided. This paper discusses the issues in existing research, and examines the need for, and concepts of, next-generation recommendation theories and techniques to address non-IIDness in recommendation. A general non-IID learning framework is proposed that captures both high-level ratings dynamics and low-level specific information on users and items and their non-IID nature.

Non-IIDness involves coupling relationships and heterogeneity in recommendation. The couplings involve subjective and objective interactions as well as explicit and implicit interactions within and between users, within and between items, and between users and items. Heterogeneity spreads from users to items, as well as to their properties. Non-IID recommendation research specifically considers the following aspects: ① low-level explicit properties of non-IID users and items involved in a recommender system; ② heterogeneity between users and between items; ③ hierarchical coupling relationships [6] within and between users and items, and between users and items; and ④ latent interactions within and between users and items, and between users and items.

Such a non-IID recommendation perspective opens paradigm-shifting opportunities and new directions for next-generation foundational research and quality recommendation. In fact, learning non-IIDness [5,6] in big data is a foundational theoretical and practical challenge in data science and big data analytics [3,13–15], which has not been paid much attention in relevant communities including computing, informatics, and statistics, because existing analytics and learning theories and systems have been mainly built on the IID assumption. The discussions about non-IID recommendation theories and systems in this work will hopefully inspire fundamental research and promising outcomes in other analytical, learning, and information-processing areas.

This paper is organized as follows. First, Section 2 discusses the intrinsic nature of recommendation problems. Section 3 presents the concept of recommendation non-IIDness. Section 4 summarizes the main issues, with a particular focus on the IID assumption that is associated with existing recommender systems and theories. Section 5 outlines the paradigm shift of recommendation research in terms of the features and generations of recommendation research. Section 6 introduces a non-IID recommendation framework and the non-IID recommendation statement. Section 7 summarizes some preliminary case studies of non-IID recommendation methods. Prospects regarding non-IID recommendation are outlined in Section 8, followed by conclusions in Section 9.

## 2. Nature of recommendation

This paper combines the multiple sources of information related to recommendation in terms of the example shown in Fig. 1. A four-table view of recommendation is built, which consists of the following four spaces of information:

- The rating information in Table A in Fig. 1. This consists of all ratings by users on items, and embeds user rating behaviors and preferences. Table A reflects the subjective information and outcomes in recommendation.
- The user information in Table B. This reflects user characteristics, properties, and relationships that drive their rating behaviors and preferences. Table B consists of objective factors of users.
- The item information in Table C. This demonstrates item

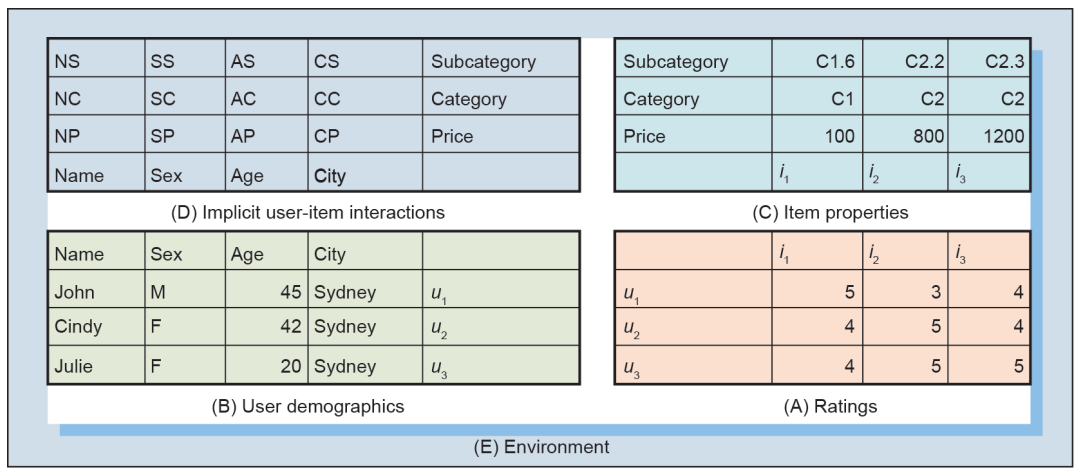

**Fig. 1.** A systematic view of recommendation.

characteristics, properties, and relationships that attract and affect users' preferences and rating behaviors. Table C consists of objective factors of items.
- The hidden user-item interactions in Table D. This is different from sources A–C, which are explicit, whereas source D is implicit, reflecting the interactions between users and items in terms of their properties. Table D consists of both objective and subjective information that connects users with items.

The above systematic view (Fig. 1) does not consider ① the environment E (shown in the outer panel), which may need to be considered in some recommendation research; or ② the interactions between Tables A–D and the environment E. A focus on this four-table view is necessary here in order to discuss the nature of recommendation. The above four-table view of recommendation discloses significant information and insights about:

(1) The specific users and items that are involved in a specific recommendation task. The user and item data show their specific properties, characteristics, and interactions. Users and items are specific and heterogeneous, since their respective features, feature values, and objects are essentially different. Users and items are personalized, holding their local characteristics and personalized preferences and interactions.

(2) How they may interact with each other and affect one another's rating behaviors and preferences. Users are coupled [6] with each other and more or less affect each other for one or multiple reasons or aspects, in the same way as items. In addition to such group or collective (global) influence, specific interactions between users and items are also deeply affected by respective individual users and items. It is important to model the balance between global and local characteristics, interactions, and preferences.

(3) How users and items affect each other. This shows that a user likes a product (embodied through ratings), probably because of specific "couplings" between the user's characteristics and the corresponding items, although such couplings are sophisticated and implicit.

(4) How objective and subjective information jointly contribute to a final decision through ratings. By connecting the subjective ratings in Table A with the objective user/item information in Tables B and C, as well as the subjective/objective user-item interactions in Table D, we are able to see the full picture of connections, drivers, and dynamics of recommendation, which is usually reflected in the ratings.

This four-table view is fundamentally different from the state-of-the-art view of recommendation, which has been built on three tables (Tables A–C) only. In addition to ratings, partial or specific user/item information is often involved. The hidden user-item interactions in Table D have not been explored in the relevant community, although they are very important for a deep understanding of how and why ratings are generated. The interactions in Table D are not as visible as those in other tables, but they incorporate implicit relations (which here we call coupling [6]) between a user attribute and an item property. (See more discussions in Section 6.5.)

The complicated implicit relationships in Table D (e.g., CP represents complex interactions between an item's price and a user's city) may be driven by underlying demographic, behavioral, social, economic, or cultural aspects, which are embodied by user and item properties and their couplings. The heterogeneity between users and between items drives the personalized rating behaviors. Therefore, there is strong non-IIDness in the four tables, which serves as the underlying driver of the ratings and preferences in Table A.

In fact, the proposed four-table view in Fig. 1 not only presents a general and unified data structure for building a comprehensive understanding of the recommendation problem; it also discloses that both explicit and implicit non-IIDness are driving factors of a recommender system.

## 3. Non-IIDness in recommendation

Following the proposed four-table view of recommendation, this section further discusses the intrinsic non-IIDness [5] in recommendation problems. In reality, any recommended items and users are non-IID; that is, there are various hierarchical coupling relationships within and between users and items, and heterogeneity exists between users and between items. Below, this section discusses these two aspects embedded in recommendation.

### 3.1. Heterogeneity

Heterogeneity is built into different aspects in the four-table view (Fig. 1). Neither users nor items are identically distributed. The following list explores scenarios of heterogeneity of users and items, and the heterogeneity between users and items.
- Heterogeneous users. Each user shares his/her own attributes, characteristics, preferences, behaviors, and intent in the ratings. Simply treating all users as identically distributed is too simple in understanding the personalized characteristics of each user, and his/her personalized demand and intent in recommendation.
- Heterogeneous items. One item is different from another in

terms of type, attributes, categories, domains, and so forth. Specific item characteristics form an attraction to different users and user ratings.

- Heterogeneous user/item attributes. Each user attribute and each item attribute is different. Each user attribute describes one respective aspect of user demographics, characteristics, group, preference, behavior, and intent. Similarly, each item attribute draws a picture of a respective aspect of item categories, types, characteristics, domains, and so on. Each user/item attribute is not identically distributed; it follows its own distribution, and thus needs to be handled accordingly.
- Heterogeneity between users and items. Users are very different from items, and cannot be assumed to follow the same distributions. Assuming that they adopt similar latent matrices or that they can be modeled in the same way is too simplistic for capturing the specific features of users and items.

If the above discussions about heterogeneity need to be considered in recommendation, many existing methods, including those based on MF, may fail to produce meaningful outcomes, and may even produce misleading recommendations. It would also not be possible to provide truly personalized recommendations when the personalized characteristics are ignored in the modeling.

*3.2. Couplings*

Considering heterogeneity is one step forward in recommendation research, although it does not capture the full picture of characteristics and complexities in recommender systems. Another important matter is to capture the explicit and implicit—which are often hierarchical—coupling relationships. Here, couplings refer to any relationships or interactions that connect two or more aspects (which could be between inputs or between inputs and outputs) [6].

As shown in Fig. 1, there are different couplings embedded in the proposed four-table view. Couplings in recommendation problems represent implicit or explicit connections between users, between items, and between users and items, for any reason or in any aspect. Further explanation is given below and illustrated in Fig. 2.

- User-user couplings. These refer to the couplings both within and between users in Table B in Fig. 2, which are further embodied through: ① intra-user attribute couplings, showing the relationships between the values of a user attribute, such as couplings between user preferences, groups, domains, behaviors, or social relationships; ② inter-user attribute couplings, showing the connections between user attributes, such as user ages and their positions; and ③ user couplings between users or between user groups.
- Item-item couplings. These are similar to user-user couplings, being couplings within and between items in Table C that consist of: ① intra-item attribute couplings, ② inter-item attribute couplings, and ③ item couplings between items or between item categories/domains.
- User-item couplings. These refer to the couplings within and between user-item pairs or clusters, which are embodied through the following aspects: ① explicit user-item couplings indicated by user ratings on items and user comments on items in Table A; and ② implicit user-item couplings, showing the influence of, or connections between, a user's attributes and the user-rated item attributes, as shown in Table D.

In addition to the different types (aspects) [6] of couplings discussed above, couplings are often presented in terms of certain hierarchies. As illustrated in Fig. 2, couplings exist in attribute values, attributes (for both users and items), objects (users and items), and object groups (user groups or item categories). In particular, user-item couplings may appear on different levels, say from couplings within the rating table (Table A) to those between rating tables of various user groups, and from couplings between a user attribute and an item attribute to those between a user attribute matrix and an item attribute matrix.

## 4. Issues with existing recommender systems

Building on the discussions about the nature of recommendation, this section discusses the issues in existing recommendation research and the IID assumption usually made in classic recommendation theories and systems.

*4.1. Related recommendation research*

Existing recommendation algorithms and methods may be broadly categorized into four families: collaborative filtering (CF), content-based filtering (CBF), hybrid approaches, and problem-specific research.

A CF algorithm [16] predicts the ratings based on a user's own and/or other users' ratings. When other users' behaviors are involved, a CF algorithm incorporates the behavior and preferences of a user group similar to those of a particular user or neighborhood [17]. However, algorithms are on item-based CF [18,19].

CBF involves user comments on items, content about items,

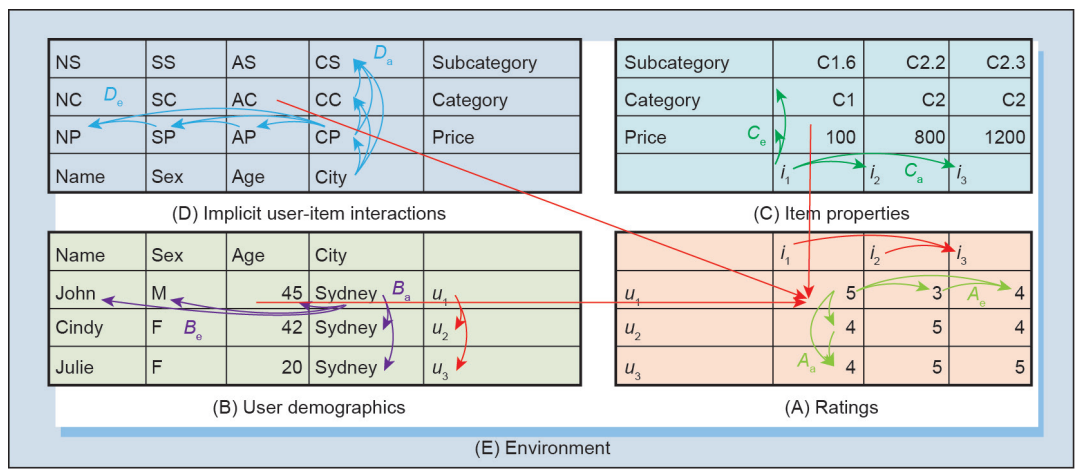

**Fig. 2.** Non-IIDness in recommendation.

and material read by users about items. This additional information is incorporated into the rating estimation.

A number of hybrid approaches exist that integrate CF and CBF in a variety of ways in parallel or serially. The hybridization may apply CF and CBF separately, followed by a merging of their results; or, CBF may first be used to identify users with similar sentiments, after which CF is employed to make recommendations on ratings. By incorporating specific factors into models such as CF, different methods have been proposed, such as social relationship-based methods [20,21].

The concept of collective matrix factorization [22] was proposed in order to factor each relation matrix with a generalized-linear link function, and, whenever an entity type is involved in more than one relationship, to tie factors of different models together. This approach captures simple and mainly latent relations between factors in observations through matrix factoring relevant relations and matrices, but it fails to capture the low-level data characteristics and complexities.

Corresponding modeling tools and evaluation metrics are proposed or used to measure recommendation performance. For example, Pearson correlation is widely used in CF, and clustering is used to group similar items or users. Typical data mining and machine learning approaches including $K$-means, $K$-modes, fuzzy $C$-means, MF, adaptive resonance theory, probabilistic clustering (expectation-maximization), Bayesian belief nets, Markov chains, and Rocchio classification are used in recommendation.

In recent years, specific recommendation problems such as cold-start, cross-domain, group-based, and shilling attack-oriented recommendations have attracted increasing attention. For example, group recommendation, which focuses on grouping user behaviors or preferences, has attracted a lot of attention [23]. Cross-domain recommendation [24,25] learns insights from user ratings on items in one domain to inform the ratings on items from other categories. Some works [26] focus on specific data and representation structures and methods for making recommendations.

More recently, following our efforts in promoting the research of non-IIDness learning [5] and non-IID recommendation [12], an emerging research direction in recommendation [6] has focused on modeling couplings within and between users, within and between items, and between users and items, and combining them with ratings. Accordingly, some preliminary work has been done, including recommendation based on item couplings [11], coupled items-based CF, and user group-based recommendation by incorporating item couplings [27]. This paper builds on these initiatives and our brief discussions in Ref. [12] to provide a comprehensive and systematic framework and discussion about non-IID recommendation learning.

### 4.2. IID assumption in classic recommendation theories

Most of existing recommendation theories and systems only or mainly involve the rating information in Table A, and focus on learning user preferences on items in terms of ratings. Typical recommendation algorithms, including those for CF and MF, often ignore the underlying reasons driving specific user preferences [18,28,29], which are drawn from user and item information in Tables B and C. They basically treat users and items as IID, and do not consider who the users are and what products they rate. To the best of this author's knowledge, no existing work has comprehensively incorporated couplings within and between ratings, users and items, and between users and items, as shown in Fig. 2.

The underlying IID assumption behind typical algorithms for CF and MF [30,31] are now analyzed.

Basic CF builds a process of filtering that involves collaborations between objects in terms of either a user-based CF or an item-based CF. A user-based CF assumes that user $b$ is likely to follow user $a$'s ratings on another item if $a$ shares the same rating as $b$ on one item. Similarly, an item-based CF assumes that users who are interested in item $x$ are also interested in item $y$. Although different CF variants have been proposed in order to address respective issues, the underlying assumption in the original memory-based CF algorithm is analyzed here [28]. Eq. (1) estimates the predicted rating $r_{a,j}$ by user $a$ based on user $i$'s rating $r_{i,j}$ on item $j$ and the mean rating $\overline{r}_i$ (where $w(a, i)$ measures the weight of the similarity between users $i$ and $a$).

$$r_{a,j} = \overline{r}_a + k \sum_{i=1}^{n} w(a,i)(r_{i,j} - \overline{r}_i) \quad (1)$$

where, $p_{a,j}$ assumes that there is only a weak correlation between users $i$ and $a$. This basic CF makes the following assumption: ① User $i$ rates all items independently, by treating items as IID and ignoring the connection between ratings $r_{i,j_1}$ and $r_{i,j_2}$ for two items $j_1$ and $j_2$. ② Two users $i_1$ and $i_2$ rate items independently, by treating users as IID, the CF overlooks the connections between two users $i_1$ and $i_2$ and their ratings $r_{i_1,j}$ and $r_{i_2,j}$. ③ User ratings on items do not affect each other; that is, they overlook the connections between users and items. ④ The algorithm does not involve user properties and item properties; rather, it uses only the rating information.

Let us further analyze the assumption in the MF approach. An MF algorithm assumes that ratings $R$ are the approximate factorization of two matrices $P$ and $Q$ representing the latent variable matrices of users and items, respectively. Accordingly, the estimation of a rating $\hat{r}_{i,j}$ of an item $j$ by a user $i$ can be made as follows:

$$\hat{r}_{i,j} = \sum_{i=1}^{k} p_{ik} q_{kj} \quad (2)$$

where, two vectors $p_{ik}$ and $q_{kj}$ capture the latent variables of user $i$ on item $j$, and the rating is affected by similar users $k$.

This basic MF makes the following assumptions: ① The rating estimation does not rely on either user or item properties. ② It assumes that the rating dynamics are driven by the user and item latent factors, and thus ignores the couplings (influence) and heterogeneity between users, between items, and between users and items.

The above analysis of underlying CF and MF principles shows that both CF and MF assume that users and items are IID, and lack the involvement of driving factors of user and item attributes in the rating estimation. Despite the diversified variants of CF and MF methods that have been proposed, if the non-IID information in Fig. 1 is not involved, it is understandable that the recommendations made may not address personalized preferences.

## 5. Paradigm shift of recommendation research

This section discusses the features of existing recommendation research, and proposes a view of four generations of recommendation research.

### 5.1. Features of recommendation research

Although the evolution of recommender systems has experienced some important periods, increasing attention is being paid to this area. Building on the discussions of existing recommendation research in Section 4, the list below summarizes typical features of state-of-the-art recommendation research, which:
- Assumes users and items are IID;
- Usually places foci on observable factors and aspects;
- Involves latent variables while ignoring explicit user/item

variables, or vice versa;
- Ignores or simplifies the interactions between explicit and implicit variables of users and items;
- Lacks deep explorations of subjective factors, and the integration of subjective and objective factors; and
- Lacks deep explorations of core driving forces and implicit interactions within and between users, within and between items, and between users and items.

*5.2. Categorization of recommendation research*

Different views exist on how to categorize recommendation research. Representative surveys on recommender systems present the following pictures about research on recommendation from different perspectives and foci of interest.
- An approach categorization of hybrid Web systems is shown in Ref. [32], consisting of four classes of recommender systems based on knowledge sources—CF, content-based, demographic, and knowledge-based systems—and 53 possible hybrid methods based on the workable combinations of seven hybridization strategies: weighted, mixed, switching, feature combination, cascade, feature augmentation, and meta-level [33] with the above four classes.
- A categorization of recommendation techniques is provided in terms of similarity, dimensionality reduction, diffusion (spreading), social filtering, meta approaches, and performance evaluation in Ref. [34].
- CF recommender systems are reviewed in Ref. [35] in terms of the evolution from algorithms to questions around the user experience with the recommender systems, issues and open problems about quality, hidden dangers, and user control.
- A recommender system taxonomy is provided in Ref. [36], which consists of four levels: memory-based (ratings), content-based (user/item features, corresponding to the traditional Web), social-based (relationship and trust, corresponding to the social Web), and context-based (user/item locations, corresponding to the Internet of Things) levels with input of both implicit and explicit data as well as user and item data.
- The relevant literature categorization and evolution from 2001 to 2010 were summarized and analyzed in Ref. [37], which categorizes them into eight application fields (books, documents, images, movie, music, shopping, TV programs, and others) and eight data mining techniques (association rule, clustering, decision tree, *k*-nearest neighbor, link analysis, neural network, regression, and other heuristic methods).
- In addition to the approach categorization in Ref. [32] cited above, Ref. [2], an edited book, collected 28 chapters that are grouped into four parts: recommendation techniques, recommender systems evaluation, human-computer interaction, and advanced topics. No informative category of recommendation research is provided in this most recent handbook.

In the literature, the following seven major categories of recommendation techniques have been focused on:
- Memory-based recommendation, which mainly focuses on rating estimation by explicit ratings from users to items or implicit valuations of items [36] by typical models such as MF and value decomposition;
- Collaborative filtering (CF), which mainly considers "user-to-user correlations" and user (or item) neighbor relationship in the user information in Table B in Fig. 1, corresponding to similar users- or items-based recommendation;
- User profiling and modeling-based recommendation, which mainly considers user demographic information in order to generate similar users w.r.t. similar demographic information for so-called "personalized" recommendation, essentially focusing on the specific user information in Table B;
- Content-based recommendation, which mainly involves item-keyword, description, and semantic indexing in item information Table C, corresponding to item preference-based recommendation;
- Group-based recommendation, which involves the social relationships and friendship in Table B in order to recommend items to associated user groups or suggest item categories to a group of users;
- Knowledge-based recommendation, which mainly involves ① domain knowledge to measure how certain item features meet users' needs and preferences and how an item meets a user's preference, such as case-based recommendations by learning relations between specific user attributes and item attributes in Table D, and ② constraint-based recommendations by applying predefined rules to associate user requirements with item attributes in Table D; and
- Hybrid recommendation, which integrates the above approaches, such as integrating CF with content-based recommendation.

The above categorization mixes information source-driven perspectives (most of them are information-driven) with function- and purpose-based approaches. They do not directly address critical challenges (such as sparsity and the shilling-attack effect) and they miss some important areas (such as the visualization and explorations in Table D).

*5.3. Taxonomy of recommendation research*

A taxonomy of recommendation research is created in Fig. 3, which consists of seven layers: application, source, goal, challenge, technique, deliverable, and evaluation of recommendation.
- Application. This refers to domain problems and applications of recommendation; recommended products, services, and channels; and so forth. Typical applications of recommendation include: mobile applications and services; social media and network services; online business and services including shopping and news; entertainment services; food and beverage services; workflow and policy suggestions; health and medical service recommendations; traveling and tourism services; marketing and customer care; business and industry services; manufacturing optimization; logistic and transport services; and digital life, including virtual reality and animating services, and living service.
- Source. This refers to data sources that may be involved in recommendation and that consist of core data and ancillary data, which may be subjective and objective, implicit and explicit. Core data includes goals and expectations, ratings, user information, item information, and user-item interaction data. Ancillary data may consist of feedback data, environmental (contextual) data, external data, domain knowledge, system data, and information from the Internet.
- Goal. This refers to the purposes of recommendation. Both business and technical goals may be associated with recommendations. From the business perspective, recommendations may be used to improve marketing and sales, customer relationship and user experience, service objectives, economic and financial goals, human-computer interactions, and website and interface design, and to suggest new business opportunities (e.g., new users, novel products, new services). Technical objectives of recommendation may focus on enhancing rating prediction, cost-effectiveness, optimization, novelty, diversity, predictability, robustness, trust, risk management, and actionability of suggestions.

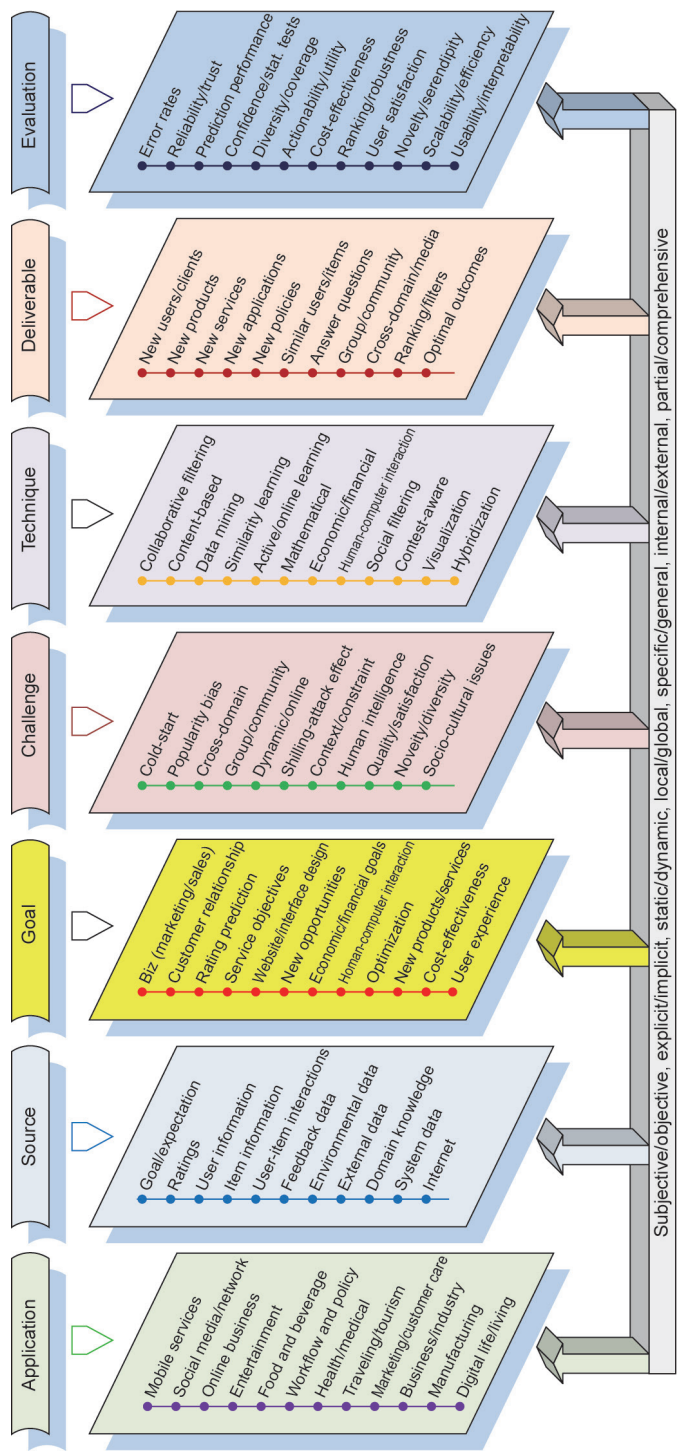

Fig. 3. The multilayer model of recommendation research.

- Challenge. This is related to the characteristics and complexities of recommendation sources (novelty, diversity, cross-domain, group and community focus, dynamic and online nature), user behavior and satisfaction (cold-start, popularity bias, shilling-attack effect, personalized satisfaction, human intelligence), environment (context, constraint, sociocultural issues), infrastructure (scalability, efficiency), performance (quality, accuracy, error rate, usability, utility, irrelevance, actionability), and so forth.
- Technique. Interdisciplinary approaches and techniques have been involved in recommendation research in terms of recommendation engine, infrastructure, algorithms, deliverables, and performance enhancement. Typical techniques include CF, content-based recommendation, data mining and machine learning methods, mathematical and statistical methods, similarity learning, active and online learning methods, economic and financial models, social science methods, context-aware techniques, visualization, and hybridization of various methods.
- Deliverable. The output of recommendation is driven by recommendation goals and techniques that are conditional on data and challenge understanding. Possible deliverables from recommendation may include suggesting similar users or products, new users, products and services, and new applications and new policies, answering or asking questions, suggesting group and community-oriented and cross-domain cross-media opportunities and experience, offering ranking and filtering suggestions, and generating optimal outcomes.
- Evaluation. Business and technical perspectives exist to evaluate the performance of recommendation. Business-wise indicators may include user satisfaction, novelty and diversity, coverage, business utility, interaction usability, and interpretability. Technical indicators may consist of improved error rates, prediction performance, reliability, robustness, serendipity, trust, confidence and statistical test performance, actionability, efficiency, and scalability.

A valid recommender system must maintain balanced interactions between the above layers. This balance involves subjective versus objective, implicit versus explicit, local versus global, specific versus general, static versus dynamic, internal versus external, and partial versus comprehensive aspects of the seven layers.

### 5.4. Generations of recommendation research

This section categorizes the research on recommendation into four major generations (Fig. 4):
- First generation (1st G): rating-based recommendation research;
- Second generation: (2nd G) user/item-based recommendation research;
- Third generation (3rd G): cross-user/item recommendation research; and
- Fourth generation (4th G): non-IID recommendation research.

The first generation mainly involves rating-based recommendation research, which corresponds to modeling and estimating the rating dynamics in the rating table (see Fig. 1) by either directly simulating the rating dynamics (such as by MF) or considering similar rating behavior and preferences (such as classic CF). Memory-based methods and specific rating characteristics are focused on in some research works, such as modeling sparse ratings, cold-start ratings, and ratings with the shilling-attack effect. At this stage, the rating information in Table A in Fig. 1 is mainly relied on in the relevant modeling.

The second generation is on user/item-based recommendation research, which corresponds to modeling rating dynamics, making user-based and item-based recommendations, and building content-based models by incorporating the specific user or item information in Tables B or C in Fig. 1. Typical examples include involving social relationships and filters between users, different categories, or subcategories of items (so-called cross-domain or hierarchical recommendation); clustering users in terms of rating behaviors or preferences; or clustering items for recommendation (so-called group-based recommendation). By involving user and item information in rating estimation and user/item recommendations, typical challenges including cold-start, sparse rating, and shilling attack are further explored, and can also be modeled by connecting to other issues such as cross-domain and group-based recommendations.

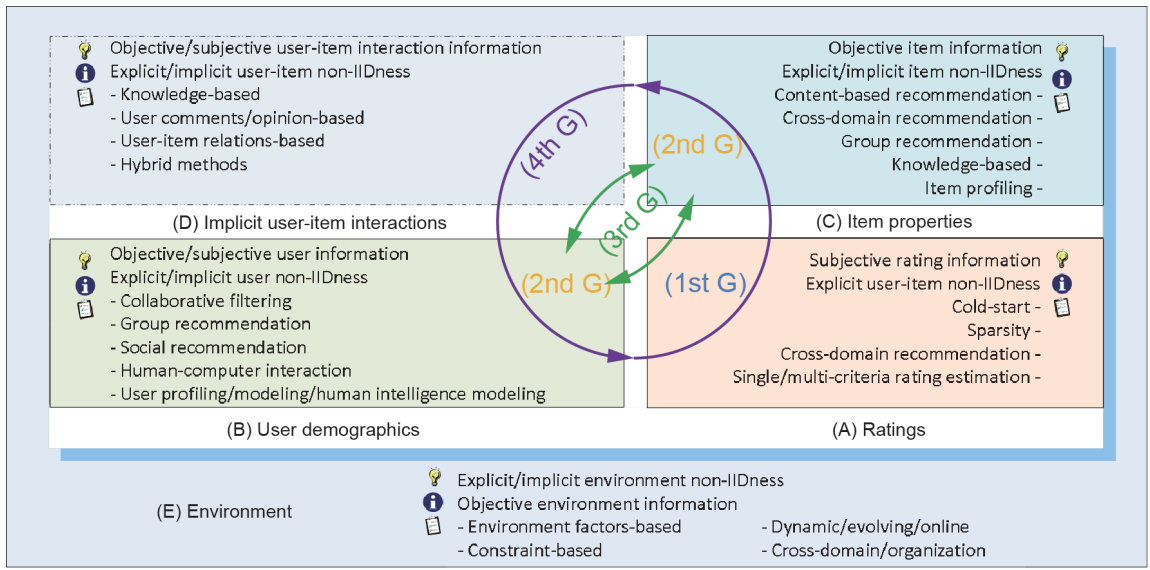

**Fig. 4.** The four generations of recommendation research.

The third generation is on cross-user/item recommendation research, which corresponds to modeling ratings and making user/item recommendations by involving the specific interaction information between users and items in Tables B and C in Fig. 1, such as user comments on products, and associations between user preferences and specific product types or characteristics. Some existing content-based modeling works fall in this category, which involves both user and item information as well as users' comments, sentiments, and opinions on items.

In existing literature, research related to the above generations makes the assumption that users and items/products are IID, and does not consider the value-to-object non-IID characteristics within and between users, products, and between users and products [38,39]. Increasing attention has been paid to learning latent variables in ratings, such as by MF-based approaches. When user and product information is incorporated, the heterogeneity and coupling relationships [5] are usually ignored.

The fourth generation is on non-IID recommendation research, which corresponds to modeling and synergizing the implicit/explicit and subjective/objective non-IIDness within and between users (in Table B), products (in Table C), and between users and products (in Tables A and D in Fig. 1). At this stage, we assume that users and products are non-IID and that they need to be considered at different levels from value to attribute and object, as well as in terms of the interactions between user attributes and product attributes. The main discussions in this paper are formed by fourth generation research, based on the systematic view in Fig. 1, which has not been explored in the literature.

Fig. 4 further maps the systematic view in Fig. 1 to cover the four generations of research on recommendation. The fourth generation actually covers the first to third generations in the sense that ① theories and approaches in the first to third generations require an IID-to-non-IID paradigm shift; and ② non-IID recommendation must involve all four tables, Tables A–D, as well as the environment E under the non-IID assumption.

## 6. Non-IID recommendation theoretical framework

This section proposes the non-IID recommendation theoretical framework, and formalizes the non-IIDness of users, items, and explicit and implicit user-item connections.

### 6.1. The non-IID recommendation framework

To effectively capture the non-IIDness discussed above in recommendation problems, a new non-IID recommendation theoretical framework is essential for building non-IID recommendation theories and systems. The principle of this non-IID framework is to capture comprehensive heterogeneity and couplings (i.e., non-IIDness) both within and between user (item) attributes, within and between users, within and between items, and between users and items.

The objectives of this non-IID recommendation framework are to:

- Incorporate both heterogeneity and couplings, namely, non-IIDness in recommendation, into recommendation algorithms and systems;
- Capture both explicit non-IIDness, such as the user-user couplings in Table B, and implicit non-IIDness, such as the user-item couplings in Table D in Fig. 2; and
- Seize both subjective non-IIDness, such as in the ratings in Table A, and objective non-IIDness, such as that of the users and items in Tables B and C.

The non-IID recommendation theoretical framework is illustrated in Fig. 5. It considers recommendation non-IIDness as in-built in the four-table view.

- **User non-IIDness.** Couplings and heterogeneity are embedded both within and between users through user attributes, attribute values, users, and user groups, illustrated in user information in Table B of Fig. 2, which can be described as a user information matrix $\mathbb{B}$, consisting of user attributes and their values on users.
- **Item non-IIDness.** Couplings and heterogeneity are inbuilt both within and between items in terms of item attribute values, attributes, items, and item categories, as shown in Table C in Fig. 2, which can be described as an item information matrix $\mathbb{C}$, consisting of item attributes and their values on items.
- **Explicit user-item non-IIDness.** Explicit non-IIDness within and between user-item connections is embodied through users' ratings on items, as shown in Table A in Fig. 2. The ratings-based user-item non-IIDness is also subjective, and is thus also called subjective user-item non-IIDness. The ratings can be represented as a rating information matrix $\mathbb{A}$, consisting of users' ratings on items.

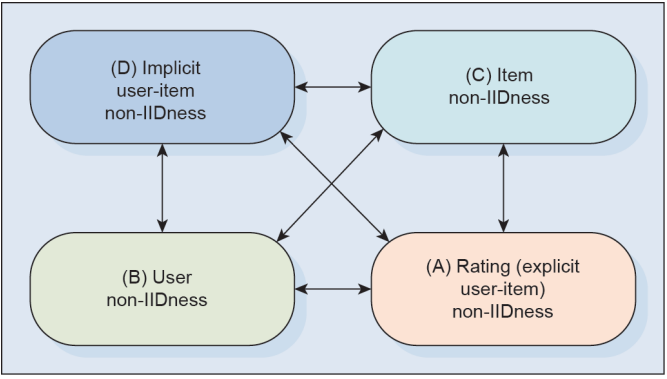

**Fig. 5.** Framework for non-IID recommendation.

- Implicit user-item non-IIDness. Implicit non-IIDness exists both within and between user-item connections through their implicit attribute interactions, as shown in Table D in Fig. 2. This can be represented by an implicit user/item information matrix $\mathbb{D}$, consisting of high dimensional relations between the user matrix $\mathbb{B}$ and the item matrix $\mathbb{C}$. Building on attribute interactions and personalized characteristics, the implicit user-item non-IIDness is objective. Therefore, the implicit user-item non-IIDness is also called objective user-item non-IIDness.

The ultimate ratings in Table A reflect the joint effect of information, interactions, and their synergy in Tables B–D.

### 6.2. User non-IIDness

As shown in Table B in Fig. 2, user non-IIDness $\mathbb{RS}^B$ is embedded in the user information table. It captures interactions, connections, and influence between users in terms of users, user attribute values, user attributes, and user grouping. For example, as shown in Table B, users $u_1$ and $u_2$ are different in terms of their ages and genders, although they are also interrelated, as they both live in Sydney and are of similar ages.

Accordingly, the user information in Table B consists of:
- Intra-user non-IIDness $B_a(\cdot)$, which captures the value non-IIDness matrix in user attribute values (this is similar to the concept intra-coupled behavior in coupled behavior analysis [38]; as an example, interested readers may refer to the intra-attribute similarity between attribute values and its similarity matrix in Ref. [39]);
- Inter-user non-IIDness $B_e(\cdot)$, which represents the non-IIDness matrix for user attributes (this is similar to the concept inter-coupled behavior for coupled behavior analysis; as an example, interested readers may refer to the concept inter-attribute similarity between attributes and its similarity matrix in Ref. [39]); and
- Aggregative user non-IIDness $B(B_a(\cdot), B_e(\cdot))$, which combines intra-user non-IIDness and inter-user non-IIDness, where $B()$ represents the integration function for combining matrices $B_a(\cdot)$ and $B_e(\cdot)$ (interested readers may refer to the coupled behavior matrix representing a coupled behavior in Ref. [38], and the coupled objective similarity matrix in Ref. [39]).

The above aspects of user non-IIDness, after being integrated, form the non-IID user space $\mathbb{RS}^B$ in Table B:

$$\mathbb{RS}^B = B(B_a(\cdot), B_e(\cdot)) \qquad (3)$$

### 6.3. Item non-IIDness

The item non-IIDness $\mathbb{RS}^C$ is indicated in the item information, Table C. It is embodied through connections and influence between items, item attributes, item attribute values, and item categorization. For example, in Table C, items $i_2$ and $i_3$ share dissimilarity between prices and subcategories, although they belong to the same category C2.

Similarly, the item information, Table C, is embedded with intra-item non-IIDness $C_a(\cdot)$, inter-item non-IIDness $C_e(\cdot)$, and the aggregative non-IIDness $C(C_a(\cdot), C_e(\cdot))$, where $C()$ represents the integration function that combines $C_a(\cdot)$ and $C_e(\cdot)$.

The above aspects of item non-IIDness, after being integrated, form the non-IID item space $\mathbb{RS}^C$ in Table C:

$$\mathbb{RS}^C = C(C_a(\cdot), C_e(\cdot)) \qquad (4)$$

### 6.4. Explicit user-item non-IIDness

The user-item non-IIDness $\mathbb{RS}^A$ in the user-item interaction, Table A, reflects the explicit and subjective interactions and influence between users and items through ratings. The user-item rating non-IIDness $A(\cdot)$ in Table A can be further decomposed to user-user rating non-IIDness and item-item rating non-IIDness. For example, in Table A, users $u_2$ and $u_3$ give the same ratings on items $i_1$ and $i_2$, but different ratings on item $i_3$.

Accordingly, $A(\cdot)$ can be categorized as intra-rating non-IIDness between users $A_a(\cdot)$, inter-rating non-IIDness between items $A_e(\cdot)$, and the aggregated rating non-IIDness $A(A_a(\cdot), A_e(\cdot))$, where $A()$ represents the integration function that combines $A_a(\cdot)$ and $A_e(\cdot)$. The overall explicit non-IID user-item similarity $\mathbb{RS}^A$ can thus be described in terms of

$$\mathbb{RS}^A = A(A_a(\cdot), A_e(\cdot)) = A(A_{j_1,j_2}, A_{i_1,i_2}) \qquad (5)$$

where, the intra-rating non-IIDness $A_{j_1,j_2}$ refers to rating non-IIDness across items $j_1$ and $j_2$; the inter-rating non-IIDness $A_{i_1,i_2}$ refers to rating non-IIDness across users $i_1$ and $i_2$.

Existing research often overlooks the non-IIDness in ratings; thus, explicit non-IID user-item similarity $\mathbb{RS}^A$ is simplified as

$$\mathbb{RS}^A = A(A_{i,j}) \qquad (6)$$

where, $A_{i,j}$ represents the preference rating matrix; $A(\cdot)$ is the aggregation function.

### 6.5. Implicit user-item non-IIDness

The most interesting and complicated non-IIDness in recommendation is embedded in Table D; namely, the implicit and objective user-item non-IIDness $\mathbb{RS}^D$. It captures the implicit and objective interactions between user attributes and item attributes. It is very complicated, since there may be hierarchical non-IIDness embedded in Table D, reflecting the interactions and influence between Table B and Table C.

In Table D, a specific cell such as "$i_1 j_1$" in subscript is called a coupled user-item cell. The non-IIDness in each cell, namely $D_{i_1 j_1}$, is the product of matrix $C_a$ for a specific item attribute $q_{j_1}$ and matrix $B_a$ for a specific user attribute $p_{i_1}$. The non-IIDness within and between cells is illustrated in Table D.

The implicit non-IIDness $\mathbb{RS}^{D_{i_1 j_1}}$ of a coupled user-item cell $D_{i_1 j_1}$ may be a matrix to learn. It consists of two parts: ① the non-IIDness of a user $i_1$'s specific attribute on all items with item attribute number $j_1$, namely $D_a$ ($D_{i_1 j_{1*}}$) ($1 \le j_{1*} \le J$); and ② the non-IIDness of an item $j_1$'s specific attribute on all users with the user attribute number $i_1$, namely $D_e$ ($D_{i_{1*} j_1}$) ($1 \le i_{1*} \le I$).

For example, in Fig. 1, SP represents the implicit couplings between a user's sex and an item's price. In addition, SP may also be influenced by other couplings in Table D, such as SC, NP, and AP. For example, the three users may be from one family, in which Cindy and John may be a couple who could affect each other's ratings. Julie may be their daughter, who may not be sensitive to price but may be more influenced by her mother. John may focus on quality while Cindy may be more sensitive to price, while Julie would be likely to balance quality and price. In addition, items $i_2$ and $i_3$ may fall into the same category.

Subsequently, the implicit non-IIDness $\mathbb{RS}^{D_{i_1 j_1}}$ in a coupled user-item cell is measured by

$$\mathbb{RS}^{D_{i_1 j_1}} = RS^{D_{i_1 j_1}}\left(D_a\left(D_{i_1 j_{1*}}\right), D_e\left(D_{i_{1*} j_1}\right)\right) \quad (7)$$

Lastly, the overall implicit user-item non-IIDness $\mathbb{RS}^D$ hidden in Table D is the aggregation of all coupled user-item cell non-IIDness in Table D, which is measured by

$$\mathbb{RS}^D = RS^D\left(\mathbb{RS}^{D_{i_1 j_1}}\right) \quad (8)$$

where, $(i_1 \neq i_2) \vee (j_1 \neq j_2) \wedge (1 \leq i_1, i_2 \leq I) \wedge (1 \leq j_1, j_2 \leq J)$.

The overall implicit user-item non-IIDness in Table D consists of two parts: ① a user $i_1$'s attribute-based item non-IIDness $D_a(\cdot)$ on all item attributes representing the non-IIDness between different item attributes on a particular user attribute, and ② an item $j_1$'s attribute-based user non-IIDness $D_e(\cdot)$ on all user attributes, indicating the non-IIDness between user attributes in terms of a specific item attribute. These two parts are further coupled in terms of $RS^D$.

$$\mathbb{RS}^D = RS^D(D_a(\cdot), D_e(\cdot)) \quad (9)$$

$\mathbb{RS}^D$ cannot be captured by a simple matrix in the same way as $\mathbb{RS}^A$, $\mathbb{RS}^B$, and $\mathbb{RS}^C$. It carries much richer information than they do from Tables B and C and their interactions on different attributes and layers and in various forms. So far, no research has been conducted on modeling the non-IIDness $\mathbb{RS}^D$.

An example to show the non-IIDness in Table D involves the cell CP in Table D, which consists of $CP_a$, capturing all couplings and heterogeneity between all item prices in a user's city, and $CP_e$, calculating the couplings and heterogeneity of all user cities on the item attribute price.

### 6.6. The recommendation non-IIDness

With the above individual non-IIDness captured for users, items, and explicit and implicit user-item interactions, we first form a complete picture of all user-item non-IIDness in recommendation, through an aggregation function $RS^{A+D}(\cdot)$ in order to combine both explicit user-item non-IIDness $\mathbb{RS}^A$ and implicit user-item non-IIDness $\mathbb{RS}^D$ as follows:

$$\mathbb{RS}^{A+D} = RS^{A+D}\left(\mathbb{RS}^A, \mathbb{RS}^D\right) \\
= \sum_{i_1,i_2=1}^{I} \sum_{j_1,j_2=1}^{J} RS^{A+D}\left(A\left(A_{i_1 j_1}, RS^D\left(RS^D_{i_1 j_1}\right)\right)\right) \odot \left(A_{i_1 j_1}, RS^D_{i_1 j_1}\right) \quad (10)$$

where, $\sum_{j_1,j_2=1}^{J} RS^{A+D}(A(A_{i_1 j_1}, RS^D(RS^D_{i_1 j_1}))) \odot$ means the subsequent non-IIDness of $RS^{A+D}$ are $A_{i_1 j_1}$ coupled with $A(A_{i_1 j_1}, RS^D(RS^D_{i_1 j_1}))$ and $RS_{i_1 j_1}$ coupled with $RS^D(RS^D_{i_1 j_1})$, and so on, with non-determinism.

The complete non-IIDness in a recommendation problem is defined below.

**Definition 1 (recommendation non-IIDness).** *The complete non-IIDness $\mathbb{RS}$ in a recommender system, namely $\mathbb{RS}$, integrates the non-IIDness from four sources: user non-IIDness $\mathbb{RS}^B$ in Table B, item non-IIDness $\mathbb{RS}^C$ in Table C, explicit user-item non-IIDness $\mathbb{RS}^A$ in Table A, and implicit user-item non-IIDness $\mathbb{RS}^D$ in Table D.*

$$\mathbb{RS} = RS(\mathbb{RS}^A, \mathbb{RS}^B, \mathbb{RS}^C, \mathbb{RS}^D) \quad (11)$$

where, $RS(\cdot)$ is the integration function.

Lastly, we define non-IID recommendation.

**Definition 2 (non-IID recommendation).** *Given a recommendation problem $\mathbb{X}$ that consists of user information matrix $\mathbb{B}$, item information matrix $\mathbb{C}$, ratings $\mathbb{A}$, and environment $\mathbb{E}$, a non-IID recommendation approach:*

*(1) Learns the complete non-IIDness $\mathbb{RS}$, including learning rating non-IIDness $\mathbb{RS}^A(\mathbb{A})$, user non-IIDness $\mathbb{RS}^B(\mathbb{B})$, item non-IIDness $\mathbb{RS}^C(\mathbb{C})$, implicit user-item non-IIDness $\mathbb{RS}^D(\mathbb{B}, \mathbb{C})$, and their synthesis method $RS(\cdot)$,*

$$\mathbb{RS} = RS(\mathbb{RS}^A(\mathbb{A}), \mathbb{RS}^B(\mathbb{B}), \mathbb{RS}^C(\mathbb{C}), \mathbb{RS}^D(\mathbb{B}, \mathbb{C})) \quad (12)$$

*(2) Learns the estimate function $\hat{\mathbb{N}}()$ conditional on environment $\mathbb{E}$ to approximate the intrinsic nature $\mathbb{N}()$ of recommended problem $\mathbb{X}$ in the physical world:*

$$\hat{\mathbb{N}}(\mathbb{A}, \mathbb{B}, \mathbb{C}, \mathbb{RS}|\mathbb{E}) \approx \mathbb{N}(\mathbb{X}) \quad (13)$$

*(3) Optimizes the objective function (e.g., loss function $\mathbb{L}() \to 0$) to obtain the most appropriate estimate $\hat{\mathbb{N}}$.*

$$\mathbb{L}() = \arg\min\left(\mathbb{N}(\mathbb{X}) - \hat{\mathbb{N}}(\mathbb{A}, \mathbb{B}, \mathbb{C}, \mathbb{RS}|\mathbb{E})\right) \quad (14)$$

## 7. Case studies of non-IID recommendation

One example of non-IID recommendation is to learn the user and item coupling relationships as discussed in Sections 6.2 and 6.3. The principle of modeling user/item couplings involves:

- Learning user similarity, which is to learn the similarities between user attribute values, between user attributes, and between users, as well as to integrate the value similarity, attribute similarity, and user similarity;
- Learning item similarity, which, similar to learning user similarity, is to learn the similarities between item attribute values, between item attributes, and between items, as well as to integrate the value similarity, attribute similarity, and item similarity; and
- Integrating user/item similarity, which is to integrate user similarity with item similarity by considering hierarchical user similarity and hierarchical item similarity.

Several preliminary works have started the exploration of modeling user/item couplings on top of classic CF and, in particular, the MF model. For example:

- Coupled item similarity-based collaborative filtering [11], in which coupled item similarity was modeled in terms of the coupled object similarity in Refs. [39,40] by incorporating coupled item attribute similarity and then introducing a coupled K-modes algorithm to predict ratings.
- Coupled item similarity-based MF [27], in which coupled item similarity was learned in terms of the coupled object similarity in Refs. [39,40] and this similarity was added into the MF objective function in order to learn latent user and item matrices.
- Coupled user/item similarity-based MF [10], in which both coupled user similarity and coupled item similarity were learned in terms of the coupled object similarity in Refs. [39,40] and both similarities were incorporated into the MF objective function for optimization.

Table 1 reports results in Ref. [10] about a comparison of cou-

**Table 1**
CMF versus CF on Movielens and Bookcrossing.

| Data set | Metrics | UBCF (Improve) | IBCF (Improve) | CMF |
|---|---|---|---|---|
| Movielens | MAE | 0.9027 (0.49%) | 0.9220 (2.42%) | 0.8978 |
| | RMSE | 1.0022 (0.18%) | 1.1958 (19.54%) | 1.0004 |
| Bookcrossing | MAE | 1.8064 (33.02%) | 1.7865 (31.03%) | 1.4762 |
| | RMSE | 3.9847 (24.68%) | 3.9283 (19.04%) | 3.7379 |

CMF: coupled matrix factorization; CF: collaborative filtering; UBCF: user-based CF; IBCF: item-based CF; MAE: mean absolute error; RMSE: root mean square error.

pled matrix factorization (CMF) against two CF methods: user-based CF (UBCF, which first computes user similarity by Pearson correlation on the rating matrix, then recommends relevant items to a given user according to those users who have strong relationships) [16] and item-based CF (IBCF, which first considers item similarity by Pearson correlation on the rating matrix, then recommends relevant items that have strong relationships with a given user's items of interest) [19]. For the latent dimension 100 for CMF, the results on Movielens indicate that CMF gains 0.49% and 2.42% w.r.t. mean absolute error (MAE) and 0.18% and 19.54% w.r.t. root mean square error (RMSE).

On Bookcrossing, CMF gains 33.02% and 31.03% in respect to MAE and 24.68% and 19.04% in respect to RMSE. This result shows that CMF substantially beats UBCF and IBCF as a result of considering the user/item couplings.

The above results of CMF show that CMF complements the subjective ratings with the objective low-level user/item information to form a comprehensive understanding of specific recommendation problems. Accordingly, CMF builds both generic and specific modeling power, whereas the basic MF only captures the generic aspects. Furthermore, the outcomes reported in Refs. [10,11,27] (interested readers can find details in Refs. [10,11,27]) show that the preliminary applications of coupled user similarity and coupled item similarity in recommender systems uncover the intrinsic low-level interactions and influence between users and between items by considering low-level attribute information and relationships.

In fact, as discussed in Ref. [5], such couplings have not been considered in the classic CF and other recommendation algorithms, which also ignore the full engagement of item attributes, user attributes, and item-user interactions in terms of attributes. This explains why such algorithms do not work well for specific applications, although they do provide generic application-independent solutions.

## 8. Prospects

At this stage, the research on recommendation is experiencing significant challenges: Many classic problems have not been well solved, while fewer and fewer breakthroughs and systematic innovations have been made. It is important to scrutinize the intrinsic complexities and nature of the underlying recommendation problems. To this end, we must learn low-level data characteristics—in particular, coupling relationships and heterogeneity of recommended users and products—and thereby focus on the new generation of non-IID recommendation research.

From the above perspectives, although recommender systems have been intensively studied, there are still great opportunities to explore theoretical breakthroughs and grand challenges. A summary of five principles for non-IID recommendation research is given here.

- Principle 1. Involve low-level and hierarchical variables and value-to-object (i.e., user, item, environment) coupling relationships into model-based approaches, in order to form data + model-driven recommendation for informed recommendation.
- Principle 2. Capture explicit and implicit variables, relationships, and specific heterogeneity in recommendation modeling in order to address the comprehensive aspects, characteristics, and complexities in recommendation problems.
- Principle 3. Learn non-IIDness in terms of users, items, and user-item interactions. In addition to the coupling learning of complex interactions [6], non-IIDness learning [5] involves many attributes, types, forms, hierarchies, structures, distributions, relations, and their synergy (see more discussions in Refs. [5,6,38]), and the challenges of combining the learning of both couplings and heterogeneity.
- Principle 4. Model the user-item non-IIDness in Table D in Fig. 2. The non-IIDness in Table D has not been studied in the relevant communities. It is important to learn the implicit and sophisticated user-item interactions in Table D, since they serve as the foundational drivers of the rating behavior and preference and their dynamics. For this, we need to learn couplings across multiple matrices and hierarchical couplings across matrices.
- Principle 5. Integrate non-IIDness in all tables in Fig. 2. This requires the involvement of both subjective and objective couplings and the explicit and implicit couplings embedded in the four tables, which are presented heterogeneously.

The proposed non-IID recommendation framework in Section 6 has great potential for creating innovative theories and systems from either individual or comprehensive perspectives; addressing existing typical challenges; and offering informed, relevant, personalized, and actionable recommendations. In addition, most of the existing works [41], including those involving social relationships, cross domain, and cross group, form special cases or only address specific aspects of the proposed non-IID recommendation framework.

In particular, the following extension and instantiation opportunities of non-IID recommendation research are discussed.

- Modeling item couplings. Modeling couplings within and between items and then incorporating their similarities into existing learning models could substantially upgrade the underlying models to cater for interactions and relations between items.
- Enhancing user profiling and modeling. Modeling the entire user information for profiling and modeling [42,43] can significantly leverage the shortage of existing models, which often focus on specific user information and aspects, such as in so-called ontological user profiling, social user profiling, implicit profiling, explicit profiling, CF-based profiling, and applications of data mining and machine learning for profiling.
- Modeling social relationships. Modeling social relationships in recommendation and social media is a special case of our proposed non-IID recommender systems, when user couplings are modeled in terms of either individual aspects such as user friendship or twittering and re-twittering interactions between users, or multiple aspects such as modeling both user friendships and user profiles. In fact, many different algorithms can be proposed by modeling user couplings in terms of either one user attribute or multiple user attributes. A major difference here is to model not only intra- but also inter-user attribute couplings. For example, in Ref. [9], a sigmoid function is used to incorporate user/item couplings into MF.
- Handling popularity bias. There are often a small number of popular items versus a large number of items on long tail, which causes a sparsity and lack of coverage issue [44]. By modeling the similarity between popular and rare users/items from the non-IID perspective, the lack of sufficient ratings information can be complemented. This can create a novel perspective to build connections between ratings on popular items and ratings on rare items, on top of existing foci on dimensionality reduction and graph-based transitive relations in the data. Interested readers may refer to Ref. [45] to obtain useful clues about indirect linkage between keywords in document analysis.
- Modeling cross-domain recommendation. Cross-domain recommendation is to suggest items from another domain to us-

ers in the underlying domain [46]. When item domain information (such as categories and subcategories, product types, and usage purposes) is focused, learning item couplings is specified to learn cross-domain factors. Similar to modeling user couplings, different algorithms can be proposed to model cross-domain item couplings in terms of either individual or multiple item attributes. Both intra- and inter-item attribute similarity should be learned. For example, the work in Refs. [24,25] models cross-domain recommendation. In some works, transfer learning was used [46,47]; these actually belong to the special cases described above. When the source and the target domain are heterogeneous, the current transfer learning may not work well. Non-IID recommendation can leverage them by modeling the non-IIDness between source and target domains.

- Modeling group recommendation. This is another special case of learning user couplings by focusing on the modeling of user grouping. For example, the work in Ref. [23] models group preferences. In Ref. [27], a coupled group MF (CGMF) algorithm was proposed, which considers user grouping in social media to cater for a specific group profile in addition to incorporating the above-discussed intra- and inter-couplings. A more sophisticated issue is to model cross-group preferences and differences when a recommendation problem involves the system-of-system phenomena [48]. In this case, we need to learn the non-IIDness between groups.
- Cold-start problem. This is to predict the ratings on newly arrived items or recommend existing items to new customers [49–53]. It addresses the issue that long-tail and new users/items receive much less feedback (or none) than those that are popular, and are usually inaccurately modeled as a result. With the non-IID recommendation principle, this problem may be better addressed by modeling the non-IID user and/or item similarities, and correspondingly making recommendation to new users or items based on non-IID user/item similarities.
- Shilling attack issue. Fake ratings are given for devious purposes [54]. This may be addressed by modeling the genuine non-IIDness in users and items, and identifying the "outlying" ratings that are not consistent with the genuine user/item non-IIDness (interested readers may refer to the coupled outlier detection method in Ref. [55]).
- Context-aware recommendation. The general understanding of contextual recommendation [56] can be treated as making recommendation within certain user/item constraints or by assuming specific user/item settings in Tables B and C, which can thus be better solved from non-IID user/item perspectives by modeling the respective non-IIDness. In addition, when the environment E in Eq. (14) excludes user/item information, such as when it refers to seasonal, economic, and sociocultural factors of a recommendation problem, we need to consider the interactions between environment and users/items, by adding the fifth table E that captures the contextual information. For this situation, the recommendation theories need to address the objectives in Definition 2.
- Human-computer interactions for recommendation. This involves: ① the effective and comprehensive acquisition and application of user information available in Table B and ② the incorporation of qualitative human intelligence, as a part of the computing/decision-making body, into recommendation. Learning user non-IIDness in Table B in Fig. 1 can address the first purpose; for example, modeling user preference and opinions on items is a special case of the user-item coupling learning in Table D in Fig. 2 that focuses on understanding user comments only. The second part involves many qualitative aspects of human intelligence [48], which is not directly available in the data source but is very important for quality recommendation. Many interdisciplinary research opportunities may emerge, such as: understanding human perceptual, cognitive, psychological, and sociocultural aspects and influence in determining how and why a recommendation is chosen or not; human decision making and choice making; the representation and understanding of human personality and preference; modeling trust and privacy and their inference in a group and collective scenario; modeling real user needs and expectations; and modeling emotional, interpersonal, experiential, attitudinal, and motivational factors of individuals and groups in recommendation and decision making [2]. This research will lead to human-machine cooperative recommendation or human-centric recommender systems by human-computer interactions [57], and to the synthesis of human and machine intelligence [58].
- Crowdsourcing for recommender systems. This involves multiple roles, including task/service requesters, workers, and providers; and multiple objectives [59,60], including reward, cost-effectiveness of delivery, skill matching, overall task completion rate, and cumulative commission. Recommendation can play a role in optimizing the multiple objectives. If we can obtain the user information for different roles, task description information, and optimization objectives, this problem can then be mapped to a multi-view problem: a multi-type user information table, a task information table, and an optimization goal table, corresponding to Tables A–C in Fig. 1. Then, the proposed non-IIDness recommendation can be applied to optimizing crowdsourcing.

## 9. Conclusions

In data economy and businesses such as social media, online business, mobile services, and advertising, recommendation plays an increasingly important role. Existing recommendation theories and systems have mainly been built on the assumption that recommended items and recommendation users are IID. This work has analyzed the issues surrounding such IID theories, and has introduced non-IIDness into recommendation by considering both couplings and heterogeneity in users and items, and between users and items. A non-IID recommendation framework has been introduced to incorporate explicit and implicit, subjective and objective, and local and global non-IIDness. This non-IID framework creates extensive challenges and will result in theoretical breakthroughs and significant innovation opportunities for next-generation recommendation research and applications.

Non-IIDness learning is a grand challenge in data science and big data analytics. It raises critical issues that confront the classic theories and tools in data analytics, information processing, statistics, pattern recognition, and learning systems. This work on non-IID recommendation study will hopefully inspire similar insights into many other topics, generating a paradigm shift from IIDness learning to non-IIDness learning, for both breakthrough theories and better practices.

## References

[1] Jannach D, Zanker M, Felfernig A, Friedrich G. Recommender systems: an introduction. Cambridge: Cambridge University Press; 2010.
[2] Ricci F, Rokach L, Shapira B, Kantor PB, editors. Recommender systems handbook. 2nd ed. New York: Springer; 2015.
[3] Cao L. Data science: a comprehensive overview. Technical report. Sydney: University of Technology Sydney; 2016.
[4] McKinsey Global Institute; Manyika J, Chui M, Brown B, Bughin J, Dobbs R, Roxburgh C, et al. Big data: the next frontier for innovation, competition, and

[5] productivity. New York: McKinsey Global Institute; 2011.
[5] Cao L. Non-IIDness learning in behavioral and social data. Comput J 2014;57(9):1358–70.
[6] Cao L. Coupling learning of complex interactions. Inform Process Manag 2015;51(2):167–86.
[7] Cao L. In-depth behavior understanding and use: the behavior informatics approach. Inform Sciences 2010;180(17):3067–85.
[8] Cao L, Yu PS, editors. Behavior computing: modeling, analysis, mining and decision. London: Springer; 2012.
[9] Fu B, Xu G, Cao L, Wang Z, Wu Z. Coupling multiple views of relations for recommendation. In: Cao T, Lim EP, Zhou ZH, Ho TB, Cheung D, Motoda H, editors Advances in Knowledge Discovery and Data mining: 19th Pacific-Asia Conference, Part II; 2015 May 19–22; Ho Chi Minh City, Vietnam. Switzerland: Springer International Publishing; 2015. p. 723–43.
[10] Li T, Lu J, López LM. Preface: intelligent techniques for data science. Int J Intell Syst 2015;30(8):851–3.
[11] Yu Y, Wang C, Gao Y, Cao L, Chen X. A coupled clustering approach for items recommendation. In: Pei J, Tseng VS, Cao L, Motoda H, Xu G, editors Advances in Knowledge Discovery and Data mining: 17th Pacific-Asia Conference, Part II; 2013 Apr 14–17; Gold Coast, Australia. Heidelberg: Springer; 2013. p. 365–76.
[12] Cao L, Yu PS. Non-IID recommendation theories and systems. IEEE Intell Syst 2016;31(2):81–4.
[13] Cao L. Data science and analytics: a new era. Int J Data Sci Analyt 2016;1(1):1–2.
[14] Cao L. Data science: intrinsic challenges and directions. Technical report. Sydney: University of Technology Sydney; 2016.
[15] Cao L. Data science: nature and pitfalls. Technical report. Sydney: University of Technology Sydney; 2016.
[16] Su X, Khoshgoftaar TM. A survey of collaborative filtering techniques. Adv Artif Intell 2009;2009(4):1–19.
[17] Koren Y. Factorization meets the neighborhood: a multifaceted collaborative filtering model. In: Proceedings of the 14th ACM SIGKDD International Conference on Knowledge Discovery and Data Mining; 2008 Aug 24–27; New York, USA; 2008. p. 426–34.
[18] Sarwar B, Karypis G, Konstan J, Riedl J. Item-based collaborative filtering recommendation algorithms. In: Proceedings of the 10th International Conference on the World Wide Web; 2001 May 1–5; Hong Kong, China; 2001. p. 285–95.
[19] Deshpande M, Karypis G. Item-based top-N recommendation algorithms. ACM Trans Inform Syst 2004;22(1):143–77.
[20] Ma H, Yang H, Lyu MR, King I. SoRec: social recommendation using probabilistic matrix factorization. In: Proceedings of the 17th ACM Conference on Information and Knowledge Management; 2008 Oct 26–30; Napa Valley, CA, USA; 2008. p. 931–40.
[21] Ma H. An experimental study on implicit social recommendation. In: Proceedings of the 36th International ACM SIGIR conference on Research and Development in Information Retrieval; 2013 Jul 28–Aug 1; Dublin, Ireland; 2013. p. 73–82.
[22] Singh AP, Gordon GJ. Relational learning via collective matrix factorization. In: Proceedings of the 14th ACM SIGKDD International Conference on Knowledge Discovery and Data Mining; 2008 Aug 24–27; Las Vegas, NV, USA; 2008. p. 650–8.
[23] Hu L, Cao J, Xu G, Cao L, Gu Z, Cao W. Deep modeling of group preferences for group-based recommendation. In: Proceedings of the 28th AAAI Conference on Artificial Intelligence; 2014 Jul 27–31; Québec City, Canada; 2014. p. 1861–7.
[24] Hu L, Cao J, Xu G, Wang J, Gu Z, Cao L. Cross-domain collaborative filtering via bilinear multilevel analysis. In: Proceedings of the 23rd International Joint Conference on Artificial Intelligence; 2013 Aug 3–9; Beijing, China; 2013. p. 1–7.
[25] Hu L, Cao J, Xu G, Cao L, Gu Z, Zhu C. Personalized recommendation via cross-domain triadic factorization. In: Proceedings of the 22nd International Conference on World Wide Web; 2013 May 13–17; Rio de Janeiro, Brazil; 2013. p. 595–606.
[26] Yang X, Steck H, Liu Y. Circle-based recommendation in online social networks. In: Proceedings of the 18th ACM SIGKDD Knowledge Discovery and Data Mining; 2012 Aug 12–16; Beijing, China; 2012. p. 1267–75.
[27] Li F, Xu G, Cao L, Fan X, Niu Z. CGMF: coupled group-based matrix factorization for recommender system. In: Lin X, Manolopoulos Y, Srivastava D, Huang G, editors Web Information Systems Engineering—WISE 2013: 14th International Conference, Part I; 2013 Oct 13–15; Nanjing, China. Heidelberg: Springer. 2013. p. 189–98.
[28] Breese JS, Heckerman D, Kadie C. Empirical analysis of predictive algorithms for collaborative filtering. In: Proceedings of the 14th Conference on Uncertainty in Artificial Intelligence; 1998 Jul 24–26; Madison, WI, USA. San Francisco: Morgan Kaufmann Publishers Inc.; 1998. p. 43–52.
[29] Resnick P, Iacovou N, Suchak M, Bergstrom P, Riedl J. GroupLens: an open architecture for collaborative filtering of netnews. In: Proceedings of ACM 1994 Conference on Computer Supported Cooperative Work; 1994 Oct 22–26; Chapel Hill, NC, USA; 1994. p. 175–86.
[30] Alter O, Brown PO, Botstein D. Singular value decomposition for genome-wide expression data processing and modeling. Proc Natl Acad Sci USA 2000;97(18):10101–6.
[31] Salakhutdinov R, Mnih A. Probabilistic matrix factorization. In: Platt JC, Koller D, Singer Y, Roweis ST, editors Proceedings of the 21st Annual Conference on Neural Information Processing Systems 2007; 2007 Dec 3–6; Vancouver, Canada; 2007. p. 1257–64.
[32] Burke R. Hybrid web recommender systems. In: Brusilovsky P, Kobsa A, Nejdl W, editors The adaptive web. Heidelberg: Springer; 2007. p. 377–408.
[33] Burke R. Hybrid recommender systems: survey and experiments. User Model User-Adapt Interac 2002;12(4):331–70.
[34] Lv LL, Medo M, Yeung CH, Zhang YC, Zhang ZK, Zhou T. Recommender systems. Phys Rep 2012;519(1):1–49.
[35] Konstan JA, Riedl J. Recommender systems: from algorithms to user experience. User Model User-Adapt Interact 2012;22(1):101–23.
[36] Bobadilla J, Ortega F, Hernando A, Gutiérrez A. Recommender systems survey. Knowl-Based Syst 2013;46:109–32.
[37] Park DH, Kim HK, Choi IY, Kim JK. A literature review and classification of recommender systems research. Expert Syst Appl 2012;39(11):10059–72.
[38] Cao L, Ou Y, Yu PS. Coupled behavior analysis with applications. IEEE Trans Knowl Data Eng 2012;24(8):1378–92.
[39] Wang C, Dong X, Zhou F, Cao L, Chi CH. Coupled attribute similarity learning on categorical data. IEEE Trans Neural Netw Learn Syst 2015;26(4):781–97.
[40] Wang C, Cao L, Wang M, Li J, Wei W, Ou Y. Coupled nominal similarity in unsupervised learning. In: Proceedings of the 20th ACM Conference on Information and Knowledge Management; 2011 Oct 24–28; Glasgow, UK; 2011. p. 973–8.
[41] Chen L, Zeng W, Yuan Q. A unified framework for recommending items, groups and friends in social media environment via mutual resource fusion. Expert Syst Appl 2013;40(8):2889–903.
[42] Nadee W. Modeling user profiles for recommender systems [dissertation]. Brisbane: Queensland University of Technology; 2016.
[43] Li R, Wang S, Deng H, Wang R, Chang KCC. Towards social user profiling: unified and discriminative influence model for inferring home locations. In: Proceedings of the 18th ACM SIGKDD International Conference on Knowledge Discovery and Data Mining; 2012 Aug 12–16; Beijing, China; 2012. p. 1023–31.
[44] Popescul A, Ungar LH, Pennock DM, Lawrence S. Probabilistic models for unified collaborative and content-based recommendation in sparse-data environments. In: Proceedings of the 17th Conference in Uncertainty in Artificial Intelligence; 2001 Aug 2–5; Seattle, WA, USA. San Francisco: Morgan Kaufmann Publishers Inc.; 2001. p. 437–44.
[45] Chen Q, Hu L, Xu J, Liu W, Cao L. Document similarity analysis via involving both explicit and implicit semantic couplings. In: Proceedings of IEEE Data Science and Advanced Analytics 2015; 2015 Oct 19–21; Paris, France; 2015. p. 1–10.
[46] Jiang M, Cui P, Chen X, Wang F, Zhu W, Yang S. Social recommendation with cross-domain transferable knowledge. IEEE Trans Knowl Data Eng 2015;27(11):3084–97.
[47] Pan W, Yang Q. Transfer learning in heterogeneous collaborative filtering domains. Artif Intell 2013;197:39–55.
[48] Cao L. Metasynthetic computing and engineering of complex systems. London: Springer-Verlag; 2015.
[49] Son LH. Dealing with the new user cold-start problem in recommender systems: a comparative review. Inform Syst 2016;58:87–104.
[50] Gantner Z, Drumond L, Freudenthaler C, Rendle S, Schmidt-Thieme L. Learning attribute-to-feature mappings for cold-start recommen-dations. In: Proceedings of the 10th IEEE International Conference on Data Mining; 2010 Dec 13–17; Sydney, Australia; 2010. p. 176–85.
[51] Mirbakhsh N, Ling CX. Improving top-N recommendation for cold-start users via cross-domain information. ACM Trans Knowl Discov Data 2015;9(4):33.
[52] Lika B, Kolomvatsos K, Hadjiefthymiades S. Facing the cold start problem in recommender systems. Expert Syst Appl 2014;41(4):2065–73.
[53] Gao H, Tang J, Liu H. Addressing the cold-start problem in location recommendation using geo-social correlations. Data Min Knowl Disc 2015;29(2):299–323.
[54] Gunes I, Kaleli C, Bilge A, Polat H. Shilling attacks against recommender systems: a comprehensive survey. Artif Intell Rev 2014;42(4):767–99.
[55] Pang G, Cao L, Chen L. Outlier detection in complex categorical data by modelling the feature value couplings. In: Proceedings of the 25th International Joint Conference on Artificial Intelligence 2016; 2016 Jul 9–15; New York, NY, USA; 2016. p. 1–7.
[56] Hidasi B, Tikk D. General factorization framework for context-aware recommendations. Data Min Knowl Disc 2016;30(2):342–71.
[57] Jacko JA, editor. The human-computer interaction handbook: fundamentals, evolving technologies and emerging applications. 3rd ed. Boca Raton: CRC Press; 2006.
[58] Qian XS, Yu JY, Dai RW. A new discipline of science—the study of open complex giant system and its methodology. Chin J Syst Eng Electron 1993;4(2):2–12.
[59] Liu X, Nielek R, Adamska P, Wierzbicki A, Aberer K. Towards a highly effective and robust Web credibility evaluation system. Decis Support Syst. 2015;79:99–108.
[60] Aldhahri E, Shandilya V, Shiva S. Towards an effective crowdsourcing recommendation system: a survey of the state-of-the-art. In: Proceedings of the 2015 IEEE Symposium on Service-Oriented System Engineering; 2015 Mar 30–Apr 3; San Francisco Bay, CA, USA; 2015. p. 372–7.